\documentclass{aa}
\usepackage{graphicx}
\usepackage{natbib}
\def\bp{\object{$\beta$\,Pictoris}}
\begin{document}
\title{Dust production from collisions in extrasolar planetary
systems} \subtitle{The inner $\beta$ Pictoris Disc.}
\author{P. Th\'ebault\inst{1}, J.C. Augereau\inst{2,3} and
H.Beust\inst{4}} \institute{ Observatoire de Paris, Section de Meudon,
F-92195 Meudon Principal Cedex, France \and CEA Saclay, Centre de
l'Orme des Merisiers 91191 Gif-sur-Yvette Cedex, France \and Leiden
Observatory, PO Box 9513, 2300 Leiden, The Netherlands \and
Laboratoire d'Astrophysique de l'Observatoire de Grenoble,
Universit\'e J. Fourier, BP 53, 38041 Grenoble Cedex 9, France}
\offprints{P. Th\'ebault} \mail{philippe.thebault@obspm.fr}
\date{Received; accepted} \titlerunning{The inner $\beta$ Pictoris
Disc} \authorrunning{Th\'ebault, Augereau \& Beust}

\abstract {Dust particles observed in extrasolar planetary discs
originate from undetectable km-sized bodies but this valuable
information remains uninteresting if the theoretical link between
grains and planetesimals is not properly known. We outline in this
paper a numerical approach we developed in order to address this
issue for the case of dust producing collisional cascades.
The model is based on a particle-in-a-box method. We follow the
size distribution of particles over eight orders of magnitude in
radius taking into account fragmentation and cratering according
to different prescriptions. A very particular attention is paid to
the smallest particles, close to the radiation pressure induced cut-off size
$R_{pr}$, which are placed on highly eccentric orbits by the stellar
radiation pressure.
We applied our model to the case of the inner ($<$\,10 AU) \bp\ disc, in
order to quantitatively derive the population of progenitors needed
to produce the small amount of dust observed in this region
($\simeq\,10^{22}$\,g). Our simulations show that the collisional
cascade from kilometre-sized bodies to grains significantly departs from
the classical $dN\propto R^{-3.5}dR$ power law: the smallest particles
($R\simeq R_{pr}$) are strongly depleted while an overabundance of
grains with size $\sim 2R_{pr}$ and a drop of grains with size $\sim
100R_{pr}$ develop regardless of disc's dynamical excitation, $R_{pr}$
and initial surface density. However, the global 
dust to planetesimal mass ratio remains close to its $dN\propto R^{-3.5}dR$
value. Our rigorous approach thus confirms the depletion in mass in
the inner \bp\ disc initially inferred from questionable
assumptions. We show moreover that collisions are a sufficient source
of dust in the inner \bp\ disc. They are actually unavoidable even
when considering the alternative scenario of dust production by slow
evaporation of km-sized bodies. We obtain an upper limit of $\sim
0.1\,M_{\oplus}$ for the total disc mass below 10\,AU. This upper limit is
not consistent with the independent mass estimate
(at least $15\,M_{\oplus}$) in the frame of the
Falling Evaporating Bodies (FEB) scenario explaining
the observed transient features activity. Furthermore, we show that the mass
required to sustain the FEB activity implies a so important mass loss
that the phenomena should naturally end in less than 1\,Myr, namely in
less than one twentieth the age of the star (at least $2\,10^{7}$\,years).
In conclusion, these results might help converge towards a coherent picture
of the inner \bp\ system: a low-mass disc of collisional debris leftover
after the possible formation of planetary embryos, a result which would be
coherent with the estimated age of the system.

 } \maketitle

\section{Introduction}

\subsection{The $\beta$ Pictoris system}

The dusty and gaseous $\beta$ Pictoris disc has been intensively
studied since the first resolved image was obtained in 1984 \citep{smi84}.
This system is particularly interesting since it is still one of the best
examples of a possible young extrasolar planetary system.  It should
be stressed that considering the estimated age of the system, i.e. at
least $2\,10^{7}$\,years \citep{bar99}, $\beta$ Pictoris is no longer
in its earliest formation stage and that if planetary accretion had to
occur then it should already be finished.

The \bp\ disc has been observed in various wavelengths and has a
radial extent of at least 1500\,AU \citep{lar01}. Due to obvious
observational constraints (10 AU at $\beta$ Pic's distance correspond
to $\sim\,0.5$"), it is mostly the outer part of the disc beyond
$\sim$30 AU that has been extensively mapped and studied with the
highest spatial resolution (the reader might refer to \citet{arty97}
for a complete review.) The global picture is that of a relative
central gap in the dust density, followed by a density peak around 100
AU and a slow decrease outwards. The exact density profile remains
relatively uncertain since it strongly depends on both the assumed
size distribution for the dust population and the grain optical
properties. As a consequence the total mass of dust is also relatively
uncertain, but might be of the order of 0.3 to
0.5\,$M_{\mathrm{\oplus}}$ \citep{ligre98,arty97}. Note that "dust" mass
estimates always strongly depend on the upper size limit
considered, and that observations do not constraint very well the
abundance of the bigger dust particles where most of the mass is
supposed to be. Observations show mostly the signature of micron-sized
grains visible through scattered light \citep[e.g.][]{kal95,mou97} and
thermal emission \citep{lapa94}.  Millimetre-sized grains are detected
by millimetre wavelength photometry \citep{chi91,zuc93} and by
resolved imaging in the submillimeter domain but with a poor angular
resolution \citep{hol98}.

Actually no direct information is available for all objects bigger
than a few millimetres.  Analytical estimates have shown that the
observed dust cannot be primordial since its expected lifetime imposed
by the rate of destructive mutual collisions is much shorter than the
estimated age of the system \citep{arty97,lag00}. Thus dust must be
constantly produced within the disc.  Candidates for producing this
dust are supposingly kilometer-sized planetesimals which may generate
dust either by evaporation of volatiles \citep{ligre98,lec98} or/and
by collisional erosion \citep{arty97,mou97,aug01}. In either case,
the total mass of the disc must be dominated by these parent bodies.
\citet{arty97} estimated that if a steady state collisional law in $dN
\propto R^{-3.5}dR$ \citep{dohn69} holds from the smallest dust grains
to the biggest planetesimals, then one might expect at least
140\,$M_{\mathrm{\oplus}}$ of kilometre-sized objects. But, as noted
by the author himself, such extrapolations remain very uncertain.

Scattered light observations have also revealed several more or less
marked asymmetries in the outer disc \citep[see][for a detailed
presentation]{kal95}. Some of these asymmetries are believed to be due
to the presence of an embedded planet. It is in particular the case of
the slight warp in inclination ($\sim\,3^{o}$) of the disc's mid plane
observed up to 80 AU. This warp has been successfully interpreted by
the dynamical response of a planetesimal disc to the pull of a
Jupiter--like object located at about 10 AU from the star on a
slightly inclined orbit \citep{mou97}. To extend this scenario to
the dust disc moreover allows to reproduce large-scale vertical
asymmetries up to about 500\,AU \citep{aug01}. The planetary
hypothesis is reinforced by new asymmetries evidenced at mid-IR
wavelengths in the inner disc \citep{wah03}.

\subsection{The inner disc}

Nevertheless, these indirect effects of an hypothetic planet are
detected much further away from the star than the planet's actual
location. As indicated in the previous section, there is a strong
lack of data for the region within 10\,AU which is probably the most
interesting area in terms of presence of planets and planet formation.
Most of the informations on this region has been indirectly inferred by
fitting the Spectral Energy Distribution (SED) in the near and middle
infrared. There seems to be a general agreement on the fact that the
inner part of the disc is significantly depleted in dust, though
opinions strongly differ on the exact extension and intensity of this
depletion \citep[see][for a detailed discussion on this
topic]{ligre98}. To the present day, one of the most complete studies
remains that of \citet{ligre98}, taking into account a large set of
parameters and especially realistic grain properties (porosity, size
distributions, chemical compositions) based on observations,
laboratory experiments and dust collection into space. This work
claims that there is no more than $6\,10^{22}$\,g of dust in the
$r<40\,$AU region, as compared to $6\,10^{25}$\,g in the [40,100]\,AU
area. The main problem is that such SED fits are strongly model
dependent, and in particular that the dust surface density
distribution cannot be uniquely determined because of its coupling to
the grain size distribution and to the optical
properties. Furthermore, the \citet{ligre98} fitting has been
performed assuming that all dust is of pure comet-evaporation origin
and that its size distribution fits in situ dust observations around
the Halley comet. As will be discussed later on (section 5), this
assumption probably cannot hold for the inner Beta-Pic disc.

There is nevertheless one independent evidence for an inner dust depletion
deduced from direct observations: \citet{pan97} obtained resolved
$12\mu$m images of the $r<100\,$AU region, with a resolution of
$\sim\,$5 AU after deconvolution. They concluded that there is a
density drop of almost an order of magnitude in the innermost
$r<10\,$AU area, although a puzzling density peak seems to be observed
at 5 AU. These authors inferred a total dust mass of
$\sim\,2.4\,10^{21}\,$g for the $r<10\,$AU area.
Note that this estimate is
also strongly model dependant, though it doesn't make any assumption
concerning the mechanism producing the dust: the authors suppose a
power law for the size distribution with a change of power law index at
a given size (and thus 3 free parameters).
The \citet{pan97} mass estimate is
significantly lower than the one that can be deduced from
\citet{ligre98} for the same region, i.e.  $\sim\,2.5\,10^{22}\,$g,
especially when taking into account the fact that the upper grain size
limit of \citet{ligre98}, 0.4\,mm, is smaller than the 1\,mm limit of
\citet{pan97}. Extrapolating the \citet{ligre98} estimate up to
the 1\,mm limit leads to a total dust mass of $\simeq\,3.5\,10^{22}$\,g.
But as mentioned before, all authors agree on one core
assumption: there $is$ a dust depletion in the inner $\beta$ Pictoris
system.
\begin{table*}[!tbph]
\caption[]{\label{summary} Summary of mass estimates for
the inner 10 AU region, as derived from previous works}
\begin{center}
\begin{tabular}{l l c r}
Authors & Modeling frame & Size range & Mass
\\
\hline
& & & \\
Li \& Greenberg\,(1998) & full SED fitting &
$[0.1\mu m, 0.4$\,mm$]$ & $2.5\,10^{22}$\,g \\
\cline{2-4}
 & when extrapolated with a $R^{-3.5}$ law & $[10, 50]$\,km & $2.5\,10^{-2}\,$M$_{\oplus}$ \\
 & & & \\
\hline \\
Pantin et al.\,(1997) & inversion of mid-IR & &\\
 & surface brightness profile & $[0.1\mu m,1$\,mm$]$ & $2.4\,10^{21}$\,g \\
\cline{2-4} & when extrapolated with a $R^{-3.5}$ law  & $[10,50]$\,km &
$2\,10^{-3}\,$M$_{\oplus}$ \\ & & & \\
\hline \\
Th\'ebault \& Beust\,(2001) & FEB scenario &
$[10,50]$\,km & $15-50\,M_{\oplus}$ \\
\end{tabular}
\end{center}
\end{table*}
\subsection{Parent bodies in the inner disc}
As pointed out by \citet{arty97} for instance, the dust is not
primordial and must be constantly replenished by larger bodies. This
author favours the scenario of dust production through collisional
erosion rather than cometary evaporation, an assumption that we also
believe to be the most reasonable one for the inner disc (see
discussion in section 5.3).  It is then tempting to use this link as
an indirect way to constrain the population of parent planetesimals,
which is otherwise totally invisible to observations. The most simple
way to do this is to suppose that a collisional equilibrium $dN
\propto R^{-3.5}dR$ power law applies from the parent bodies down to
the observed micron-sized grains. This is the usual assumption commonly
made to easily derive parent bodies masses in extrasolar dust discs
(e.g. \citet{aug99} for HR 7496 A).
This would here lead to a mass of objects
in the 10 to 50 km range comprised between
$2\,10^{-3}\,M_{\mathrm{\oplus}}$ \citep[taking the][dust
density]{pan97} and $2.5\,10^{-2}\,M_{\mathrm{\oplus}}$ \citep[taking
the][estimate]{ligre98}. These values seem very low, especially
compared to the 140\,$M_{\mathrm{\oplus}}$ mass of planetesimals
estimate for the whole system \citep{arty97} which was in accordance with the
picture of an "early Solar System". This tends to reinforce the image
of a strong mass depletion in the inner disc.  Note that the concurrent
cometary evaporation scenario also gives a quantitative link between
the observed dust and the source kilometre-sized comets
\citep[e.g. Equ.2 of][]{lec98}, but this estimate does not constrain
the number of non-evaporating objects.

There is nevertheless another way to get informations on the
planetesimal population
through the study of the so called Falling Evaporating Bodies
(hereafter FEB) phenomenon.  It is indeed believed that the
evaporation of at least kilometer-sized bodies is responsible for the
transient absorption features regularly observed in various spectral
lines: CaII, MgII, FeII, etc... \citep[e.g.][]{bog91,vid94,xxii}.
Several theoretical and numerical studies have shown that these FEB
might be bodies located at the 3:1 and/or 4:1 resonances with a giant
planet on a slightly eccentric orbit located around 10 A.U. These
objects are excited on high eccentricity $e$ orbits which allow them to
pass sufficiently close to the star, i.e. less than 0.4\,AU, for
silicate to evaporate \citep[see][and references
therein]{bm00,thebeu01}. \citet{thebeu01} estimated that the number
density of planetesimals required to fit the observed rate of
absorption features would lead to a mass of $\simeq$\,15--50\,$M_{\oplus}$
objects in the 10 to 50 km range when assuming an equilibrium
differential law in $R^{-3.5}$ in the inner $<$10 AU region. This
very high estimate is close to the \citet{arty97} estimate
for the $whole$ disc and strongly exceeds, by at least a factor
$10^{3}$, from the above-mentioned much lower dust-mass-extrapolated
estimations for the inner disc.

\subsection{The need for a numerical approach}

There is thus yet no coherent picture of the inner $\beta$-Pic
system's structure, especially for the crucial link between the
observed dust and unseen bigger parent bodies.  The main reason for
this is that deriving mass estimate from a simple power law from the
micron to the kilometre might be strongly misleading.

A first argument is that a very small difference in the power law
index leads to enormous differences when extrapolating it over such a
wide size range. If $q$ is this index, then the mass ratio between 2
populations of sizes $R_{1}$ and $R_{2}$ reads $M_{1}/M_{2} =
(R_{1}/R_{2})^{(q+4)}$. As an example, the incompatibility between the
$15-50\,M_{\oplus}$ FEB mass estimate and the
$2\,10^{-3}$--$2.5\,10^{-2}\,M_{\oplus}$ extrapolated from the observed
dust density might be solved when changing the $q$ index in the later
extrapolation from -3.5 to -3.2.  But the single power law approach
raises also other problems.  Firstly, there is no reason why the upper
size limit of the collisional cascade should be 10 or
50\,km. Extrapolating a $q\,=\,-3.5$ power law up to, say, 1000\,km
instead of 50\,km, would lead to a 4.5 times superior mass of large
objects, thus reducing the magnitude of the inner mass depletion. But
then, taking the same upper limit for the rest of disc (i.e. outside
10 AU) would increase the total mass of the system to unrealistically
high values (more than 1000\,$M_{\oplus}$).  In this case two
different size distributions should hold for the inner and the outer
systems.

Secondly, it is almost certain that a single -3.5 equilibrium power
law cannot hold over such an extremely large size range. For such a
power law to apply, all particles in the system should have reached
mutual collisional equilibrium. This is perhaps not the case here,
especially for the bigger bodies which, depending on their number
density, might have low collision rates.  Furthermore, such a power
law is theoretically achieved only for an infinitely small lower size
cutoff. As shown by \citet{bag94}, any finite size cutoff will give
rise to wavy size distribution structures which can strongly differ
from the theoretical -3.5 slope.  The reason for such size
distribution waves is simple: the smallest particles will be
over-abundant since they have no smaller bodies to destroy them. This
over-abundance will give rise to an under-abundance for all bigger
bodies that might be collisionally fragmented by these minimum-sized
objects. This will in turn lead to an over-abundance of bigger bodies,
and so on.  This point is of great importance here since there is an
obvious size cutoff for our system, i.e. the smallest grains that are
not blown away by the star's radiation pressure and which are
typically micron-sized.  Apart from this cutoff effect, the smallest
grains are also expected to have a very peculiar behaviour: even if
radiation pressure is not able to remove them, it should nevertheless
place them on highly eccentric orbits, thus augmenting their impact
velocities and shattering power, but at the same time reducing their
density in the inner region since they will spend most of their orbits
very far away from the star.  The physical link between dust and
planetesimals is thus a complex one, that cannot be handled by simple
analytical power laws.

We propose here to address these problems by performing accurate
numerical simulations. A statistical particle-in-a-box code is used to
study the mutually coupled collisional evolution of a swarm of objects
ranging in size from large planetesimals down to the smallest
micron-sized grains. The code is similar to the ones developed for
asteroid populations studies but stretches down to very small dust
particles and takes into account the peculiar dynamical evolution of
micron-sized grains submitted to the star's radiation pressure. We
detail in section \ref{model} and \ref{dust} the numerical approach we
use to derive a realistic grain size distribution at a distance of a
few AU resulting from a collisional cascade in the radiative
environment of \bp. We explore the impact of the free parameters,
along with the two extreme surface densities independently deduced
from dust and gas observations of \bp, on the final size distribution
after 10\,Myr (section \ref{results}.)  We then discuss in section
\ref{discussion} the implications of our approach and show how it
helps to go towards a coherent view of the inner \bp\ disc (section
\ref{conclusion}).
\section{Numerical procedure}
\label{model}

We will here follow the classical particle in a box approach
used by models studying the asteroid belt size distribution
\citep[e.g.][]{pet93}. We consider a typical annulus of material in the
inner disc, of radius 1\,AU and located at 5\,AU from the star.  The
system is divided into $n$ boxes accounting for each particle size
within the annulus. The size increment between two adjacent bins is
$2^{1/3}$.  At each time step the evolution of the number $dN_{k}$ of
bodies of size $R_{k}$ is given by

\begin{equation}
dN_{k} = \sum_{i,j=1}^{n}\,n_{i,j,k}p_{i,j}\,N_{i}N_{j} dt
- \sum_{l=1}^{n}\,\gamma_{l,k}p_{l,k}\,N_{l}N_{k} dt
\label{dn}
\end{equation}

where $n_{i,j,k}$ is the number of
fragments injected into the $k$ bin by an impact between 2 bodies in the $i$
and $j$ bins and $p_{i,j}$ the impact rate for a pair belonging to
the same two bins. The last term of the equation accounts for the
loss of $k$ objects due to destructive collisions ($\gamma_{l,k}=1$ for a
catastrophic fragmentation impact and $0<\gamma_{l,k}<1$ (satisfying the
mass conservation condition) for a cratering event).
The temporal evolution of $N_{k}$ is then computed using a first order
eulerian code with a variable time step. 
The key information needed for estimating
$n_{i,j,k}$, $\gamma_{l,k}$ and $p_{i,j}$ is the dynamical state of the system,
which can be parameterised by the average impact velocities $\langle
dv\rangle $.

\subsection{Dynamical state of the system}

As stated in the previous section, there is yet no clear picture of
the system we intend to study. We have in particular no precise idea
of the dynamical state of the inner disc.  There is nevertheless some
indirect evidence of the dynamical state in the outer parts, given by
the observed thickness of the disc.  The disc aspect ratio in the 100
AU region is believed to be $\simeq$ 0.1 \citep{aug01}. Thus, a first
order approximation of the average inclination of the observed dust
particles would be half this value, i.e. $\simeq$0.05\,rd.
However, this value only gives very partial
information, and this for several reasons:
\begin{enumerate}
\item It is not straightforward to extrapolate it to the
inner regions of the disc, where the dynamical conditions could be
completely different. Indeed \citet{kal95} seem to observe a
significant departure from constant disc opening for $r<60$\,AU, but
such determinations should be taken with great care, since measures of
the disc's thickness become very uncertain for these inner regions.

\item This value holds for the observed micron to millimetre-sized grains
population. It is not at all certain that bigger parent bodies have
the same inclinations

\item The dynamical state of the system depends on
the inclination $and$ eccentricity distributions.
The $\langle e\rangle $ value cannot be directly deduced from
$\langle i\rangle $, at least for
the micron-sized population, where orbits might strongly depart from
the equilibrium $\langle i\rangle = \langle e\rangle /2$ equipartition
relation (points 2 and 3 will be discussed in more details in section
\ref{dust}).
\end{enumerate}
We will thus take the $\langle e\rangle $ and $\langle
i\rangle $ values of the parent bodies in the considered $r<10\,$AU
region as free parameters (see section \ref{ei}), but we will
nevertheless refer to the $\langle i\rangle = 0.05 = \langle e\rangle
/2$ case as our "nominal" case.  Note that our simulations do not
directly use the $\langle e\rangle $ and $\langle i\rangle $
parameters but the average relative velocity parameter $\langle
dv\rangle $ given by \citet{lisste93}:
\begin{equation}
\langle dv\rangle = \left( \frac{5}{4} \langle e^{2}\rangle + \langle
i^{2}\rangle \right)^{1/2}\,\langle v_{kep}\rangle
\label{dv}
\end{equation}
where $\langle v_{kep}\rangle $ is the average Keplerian velocity of
the bodies.  Furthermore, we will assume the same $\langle e\rangle
=\langle e_{0}\rangle $, $\langle i\rangle =\langle i_{0}\rangle $ and
thus $\langle dv_{i,j}\rangle =\langle dv_{0}\rangle $ for all
particles, with the important exception of the micron-sized grains
which are significantly affected by the star's radiation pressure (see
section \ref{dust} for more details).  We will also make the
simplifying assumption that our particle in a box system is not
dynamically evolving, so that $\langle dv\rangle $ remains constant
throughout the run.

\subsection{Density of objects}

As described in sections 1.2 and 1.3, there are two independent
estimates of the density of bodies in the inner disc: 1) a dust mass
(all bodies smaller than 1 mm) of $2.4\,10^{21}-3.5\,10^{22}$g derived
from fits of the observed SED 2) a mass of $15-50 M_{\oplus}$ of
planetesimals in the 10-50 km range required to sustain the FEB
activity.  As previously discussed, these 2 estimates appear totally
incompatible when assuming an equilibrium differential $R^{-3.5}$ size
distribution throughout the system, since in this case the mass of
planetesimals extrapolated from the dust estimate is only
$2\,10^{-3}-2.5\,10^{-2}\,M_{\oplus}$.  In order to check how strong an
incompatibility there really is, or if there is any incompatibility at
all, we will consider two extreme initial discs (see sections
\ref{nominal} and \ref{massive}):
\begin{itemize}
\item An initial $R^{-3.5}$ distribution extending from $R_{min}=
2^{-2/3} R_{pr}$ (i.e. 2 boxes under the $R_{pr}$ ejection size, see
section \ref{dust}) up to $R_{max} = 50$ km which is compatible with observed
dust mass estimates. We chose to take an
initial $dust$ mass of $2\,10^{22}$\,g in the whole inner disc
(i.e. $\simeq 2\,10^{21}$g in the considered 1 AU annulus around 5
AU), an intermediate value between the \citet{pan97} and
\citet{ligre98} estimates. This leads 
to a total initial mass of $1.8\,10^{25}$\,g for the
whole system.  This initial density distribution will be taken for our
"nominal" case\\
\item An initial $R^{-3.5}$ distribution extending from $R_{min}= 2^{-2/3}
R_{pr}$ up to $R_{max} = 50$ km and compatible with an average FEB
estimate of $\simeq 25 M_{\oplus}$ of 10-50 km-sized objects in the
1--10 AU region (i.e. $\simeq 2.5\,M_{\oplus}$ in the considered
annulus) , which leads to a total initial mass of $3.5\,10^{28}$\,g\\
\end{itemize}
Regardless of the initial density distribution, the number of size
boxes considered is always the same for a given $R_{pr}$, ranging from
$R_{min}= 2^{-2/3} R_{pr}$ to $R_{max}=50$km, with two adjacent boxes
separated by a factor $2^{1/3}$ in size, thus leading to a total
number of 103 boxes for the "nominal" case where $R_{pr}=5\,10^{-4}$cm
(see section \ref{dust})

\subsection{Threshold specific energy}

The core of such a code is the prescription giving $n_{i,j,k}$ for a
given $\langle dv\rangle $.  Basically, impacts can be divided into
two categories: {\rm catastrophic fragmentation} and {\rm
cratering}, depending on the value of the impacting energy as
compared to $Q_{*}$, the threshold specific energy of the bodies,
which represents their resistance to shattering and is deduced from
laboratory experiences and analytical considerations. $Q_{*}$ is by
definition the value of the specific energy $Q$ (the ratio of the
projectile kinetic energy to the target mass) when the mass of the
largest remaining fragment $M_{lf(i)}$ is equal to $0.5\,M_{i}$.

The problem is that estimations of $Q_{*}$ do strongly differ from one
author to another \citep[see figure 8 of][for an overview]{benz99}.
Basically, all authors agree on one core assumption, i.e. the response
of solid bodies to impacts is divided in two distinct regimes: the
$strength$ $regime$ for small bodies, where the object's resistance
decreases with size, and the $gravity$ $regime$ for larger objects
where resistance increases with size because of the object's
self-gravity \citep[e.g.][]{hh90}. Nevertheless, the slopes and turn
over size from one regime to another are still a great subject of
debate. We will here consider separately two different $Q_{*}$
prescriptions (see section \ref{Qstar}):
\begin{itemize}
\item the global strength + gravity regime law given in Equ.6 of
\citet{benz99} (nominal case)
\item the \citet{hh90} law for the strength regime completed by the
\citet{hol94} law for the gravity regime.
\end{itemize}

Our code calculates for every target-impactor couple $(i,j)$ the
corresponding value of $Q_{*(i,j)}$.  Let us term $F_{lf(i,j)}$ the
ratio $M_{lf(i)}/M_{i}$. From the value of $Q_{*}$, $F_{fl(i,j)}$ can
be inferred through the empirical relation \citep{fuj77}:
\begin{equation}
F_{lf(i,j)}\,=\,0.5 \left(\frac{Q_{*}M_{i}} {E_{rel}} \right)^{1.24}
\label{lf}
\end{equation}
where $E_{rel}$ is the relative kinetic energy of the system given by
$E_{rel}\,=M_{i}M_{j}dv^{2}/2(M_{i}+M_{j})$. Note that this relation
is valid only for head-on impacts and has to be corrected by taking
into account its value averaged over all impacts angles. We will here
follow \citet{pet93} and take the average value:
\begin{equation}
\overline{F_{lf(i,j)}}\,=\,3F^{2/3}_{fl(i,j)} \,-\, 2F_{fl(i,j)}
\label{lfm}
\end{equation}
Thus, Equ.\ref{lf} has to be corrected by a numerical factor
$x_{cr} = 4^{-1/1.24}=0.327$, since
$\overline{F_{lf(i,j)}}=1/2$ for $F_{lf(i,j)}=1/8$.

\subsection{Fragmentation}

Catastrophic fragmentation occurs by definition when $F_{lf(i,j)}$ is
less than $0.5$. If we suppose that the produced fragment size
distribution follows a single-exponent power law $dN\,=\,CR^{q}dR$,
then there is a unique set of values for $q$ and $C$ derived from the
value of $F_{lf(i,j)}$ and the mass conservation condition. As pointed
out in several previous studies, this single power law specification
is the easiest to handle in models but it is a strong
oversimplification. It gives rise to several problems, in particular
the possibility to get so-called "supercatastrophic" impacts where
$q\,<\,-4$, for which there is a divergence of the total mass when
taking infinitely small lower cutoff. As noted by \citet{tang99}
"...values beyond -3 for the exponent of the cumulative size
distribution cannot hold down to very small sizes, because this would
lead to unreasonably large reconstructed masses.  For this reason it
is clear that, at some value of the size, the distributions are
expected to have a definite change of slope". Note that this change of
slope between the small and large fragments domain is also supported
by experimental experiments \citep{dav90}. This problem is
particularly crucial for the present study since our size cutoff is
extremely small (see below).

As a consequence, we will here adopt 2 different power laws of index
$q_{1}$ and $q_{2}$, each holding for a different mass range and
always taken such as the small mass index $q_{2}$ is smaller than
$q_{1}$.  The main problem is to determine where the change of slope
occurs and what the difference in slope is. We shall remain careful
and keep the slope changing size $R_{s(i)}$ as well as the ratio
$q_{1i}/q_{2i}$ as free parameters that will be explored in the runs
(see section \ref{fragcrat}). Note that once $R_{s}$ and
$q_{1i}/q_{2i}$ are given, the values $q_{1i}$ and $q_{2i}$ for the
fragments produced on a target $i$ by an impactor $j$ are uniquely
determined through the set of relations:
\begin{equation}
M_{1i}\,=\,\left[\frac{b_{1i}M^{b_{1i}}_{lf(i)}} {(1-b_{1i})}
\left(M^{(1-b_{1i})}_{lf(i)}\,-\,M^{(1-b_{1i})}_{s(i)}\right)
\,+\,M_{lf(i)} \right]
\end{equation}
\begin{equation}
C_{1i}\,=\,3b_{1i}R^{3b_{1i}}_{lf(i)}
\end{equation}
\begin{equation}
C_{2i}\,=\, \frac{3b_{1i}R^{3b_{1i}}_{lf(i)}} {R^{3(b_{1i}-b_{2i})}_{s(i)}}
\end{equation}

\begin{equation}
\frac{C_{2i}}{(3-3b_{2i})}R^{-3b_{2i}}_{s(i)} \,=\,
\frac{M_{i}-M_{1i}}{M_s(i)}
\label{coef}
\end{equation}
where $b_{1i}=-\frac{1}{3}(q_{1i}+1)$ and $b_{2i}=-\frac{1}{3}(q_{2i}+1)$,
$M_{s(i)}$ is the mass of an object of size $R_{s(i)}$, $M_{1i}$ is
the total mass of fragments produced in the domain where the
$dN=C_{1i}R^{q1i}dR$ law applies, i.e. between the size of the slope
transition $R_{s(i)}$ and the size of the largest fragment
$R_{lf(i)}$, and $C_{1i}$ and $C_{2i}$ are the coefficients for the
two $dN=C_{1i}R^{q1i}dR$ and $dN=C_{2i}R^{q2i}dR$ power laws.  This
set of equation is solved numerically for each $(i,j)$ couple.

\subsection{Cratering}

For the cratering case ($F_{lf}>0.5$), we will take the simplified
prescription of \citet{pet93}, where a fixed power law index $q_{c}=-3.4$ is
considered. The total mass of craterized mass is given by

\begin{equation}
M_{cr}\,=\,\alpha E_{rel} \,\,\,\mathrm{for}\,:\,\, E_{rel}\leq\frac{M_{i}}{100\alpha}\:,
\end{equation}

\begin{equation}
M_{cr}\,=\,\frac{9x_{cr}\alpha}{100Q_{*}\alpha-x_{cr}}
E_{rel} \:+\: \frac{M_{i}}{10}
\frac{x_{cr}-10Q_{*}\alpha}{x_{cr}-100S_{*}\alpha}
\end{equation}

\begin{displaymath}
\;\;\;\mathrm{for}\,:\,\,E_{rel}>\frac{M_{i}}{100\alpha}
\end{displaymath}
where $\alpha$ is the crater excavation coefficient which depends on
the material properties. We will explore values of $\alpha$ (section
\ref{fragcrat}) ranging from $10^{-9}$ to $4.10^{-8}\,s^{2}.cm^{-2}$,
the extreme values corresponding to "hard" and "soft" material
respectively \citep[see][and references therein]{pet93}, and take
$10^{-8}\,s^{2}.cm^{-2}$ as our "nominal" value.  The mass of the
largest fragment produced by the impact is then equal to
$F_{l_{cr}}\,M_{cr}$, where $F_{l_{cr}} = 1+\frac{1}{3}(q_{c}+1)$.

\subsection{Fragment reaccumulation}

The fraction of fragmented material reaccumulated onto the parent
bodies is the result of the competing ejectas' kinetic energy and the
parent bodies' gravitational potential. We will make the simplified
assumption that all fragments produced after an $(i,j)$ impact have
the same velocity distribution \citep{stecol97}:
\begin{equation}
v_{fr}= \left(2\frac{ f_{ke} E_{rel_{(i,j)}} }{M_{i}}\right)^{1/2}
\label{fragvel}
\end{equation}
where $f_{ke}$ is the fraction of kinetic energy
that is not dissipated after an impact. We will make here the
classical assumption that $f_{ke}=0.1$ for high velocity impacts
\citep{fuj89}. The mass fraction of fragment material that escapes
the target+impactor system is given by \citep{stecol97}:
\begin{equation}
f_{esc}=0.5 \left (\frac{v_{esc}} {v_{fr}}\right)^{-1.5}
\label{fracesc}
\end{equation}
where $v_{esc}$ is the escape velocity of the colliding bodies system.

\section{The specific behaviour of the micron-sized population}
\label{dust}

The main challenge of this simulation is that we would like to study
the collisional correlation between objects ranging from the
micron--sized to the kilometre--sized domain, i.e. separated by a 8
orders of magnitude in size. Our lower cutoff is indeed the "real"
physical cutoff $R_{pr}$ imposed by the effect of the star's radiation
pressure. Radiation pressure also strongly affect particles bigger
than $R_{pr}$, but still in the same size range, by placing them on
highly eccentric orbits. These eccentricities depend on the ratio
$\beta$ between the radiation pressure force $F_{pr}$ and the
gravitational force $F_{grav}$. For a particle produced by a parent
body on a ($a_{0},e_{0}$) orbit at a distance $r_{0}$ from the star,
one gets~:
\begin{equation}
a_{k} = \frac{1-\beta} {1-2a_{0}\beta/r_{0}} \, a_{0}
\label{pra}
\end{equation}
\begin{equation}
e_{k} = \left( 1 - \frac{ (1-2a_{0}\beta/r_{0})(1-e^{2}_{0}) }
{(1-\beta)^2} \right)^{1/2}
\label{pre}
\end{equation}
where $a_{k}$ and $e_{k}$ are the produced grain's semi-major axis and
eccentricity and $a_0$ and $e_0$ the semi-major axis and eccentricity
of the parent body. From Equation \ref{pre} and with the assumption
that the planetesimals releasing dust particles by collisions are
mostly on circular orbits ($e_0\simeq 0$, $a_0\simeq r_0$), then
grains with $\beta \geq 0.5$ are ejected from the system on hyperbolic
orbits. The cutoff size $R_{pr}$ is by definition the size for which
$\beta = 0.5$ and grains with sizes $R$ larger than $R_{pr}$ are
related to $\beta$ through the relation $\beta=0.5(R_{pr}/R)$. The
blow-out size $R_{pr}$ depends on the stellar spectra and on grains
optical properties. For $\beta$ Pictoris we use a low-resolution A5V
spectra derived from Kurucz stellar models and we adopt the chemical
grain composition proposed by \citet{ligre98} \citep[we refer to the
latter paper and to][for a full discussion on chemical and optical
properties of the grains assumed here]{aug99}.
Bare compact silicate (``Si'') grains in the surroundings of $\beta$
Pictoris and smaller than $R_{pr,\,{\rm compact}}\simeq 3.5\,\mu$m
have $\beta \geq 0.5$. The same grains but coated by an organic
refractory (``or'') mantle in a Si:or volume ratio of 1:2 as proposed
by \citet{ligre98} are ejected from the system on unbound orbits if
they are smaller than $R_{pr,\,{\rm compact}}\simeq
2.5\,\mu$m. Actually the main uncertainty on $R_{pr}$ relies on the
grain porosity $P$. The
porosity affects the optical properties of the grains and consequently
$F_{rad}$. But actually the $\beta$ ratio more dramatically depends on
$P$ through the grain density in $F_{grav}$ especially for large
porosities. The density of porous grains is related to the density of
the same but compact grain by the simple relation~: $\rho_{\rm porous}
= (1-P)\rho_{\rm compact}$ which implies for large porosities~:
$R_{pr,\,{\rm porous}} \simeq (1-P)^{-1} R_{pr,\,{\rm compact}}$.
From SED fitting, \cite{ligre98} constrained $P$
in the narrow range [0.95, 0.975]. But these values are obtained when
assuming that all grains are of $cometary$ origin, an assumption that
we believe might not hold for inner \bp\ disc (see the more complete
discussion in section 5.3.) In the present paper we keep $P$ has a
free parameter with $R_{pr}=5\,\mu$m (i.e. $P \simeq 0.5$)
taken as our reference nominal case (see section \ref{porosite}).

The radiation pressure induced eccentricity expressed by Equation
\ref{pre} is significant, say $\geq 0.1$, for all particles comprised
between $R_{pr}$ and $\sim5R_{pr}$. Thus all objects in this size
range will have orbital characteristics that depart from the general
average values defined in section 2.1. This will significantly affect
the collision rates and physical outcomes for impacts involving these
small grains.  For these impacts, Equ.\ref{dv} is no longer valid in
its simple form and $\langle dv_{i,j}\rangle $ will be numerically
estimated. To do this, we use a simplified version of a deterministic
collisional model \citep[][and references therein]{themascho02} to
derive average $\langle dv\rangle $ between a population of targets
having the nominal orbital characteristics as defined in section 2.1
and a population of impactors with a given $\beta_{k}$ (i.e. $a_{k}$
and $e_{k}$), all randomly distributed within the 1--10 AU region.

Another major consequence of these radiation-induced high $e_{k}$ is
that small grains will spend a significant fraction of their orbits
$outside$ the inner disc.  Thus, at a given moment, only a fraction
$f_{i(k)}\,N_{k}$ of these bodies will actually be present in the
considered system.  These $\langle f_{i(k)}\rangle $ are numerically
estimated with a simple code randomly spreading 10000 test particles
of a given $\beta_{k}$ uniformly produced in the 1--10 AU region
(table \ref{tabf}).

\begin{table}
\caption[]{numerical estimate of $\langle f_{i(k)}\rangle $ in the 5
AU region, for a swarm of grains produced in the whole 1--10 AU
region, as a function of $\beta_k$ (see text for details). Note that
for low $\beta$, $e_{k}$ might become lower than the average
eccentricity for the parent bodies in the disc (see section 2.1). In
such a case, the radiation pressure effect is neglected for all
$e_{k}<\langle e_{0}\rangle $ and the values of $\langle
f_{i(k)}\rangle $ rescaled so that $\langle f_{i(k)}\rangle =1$ for
$e_{k}=\langle e_{0}\rangle $.  }
\label{tabf}
\begin{center}
\begin{tabular*}{\columnwidth}{cccc}
\hline \multicolumn{1}{c}{$\beta_k$} & \multicolumn{1}{c}{$e_{k}$} &
\multicolumn{1}{c}{$\langle f_{i(k)}\rangle $}\\
0.49 & 0.96 & 0.038\\
0.39 & 0.63 & 0.344\\
0.31 & 0.45 & 0.521\\
0.24 & 0.32 & 0.665\\
0.19 & 0.24 & 0.753\\
0.15 & 0.18 & 0.812\\
0.12 & 0.14 & 0.849\\
0.10 & 0.11 & 0.887\\
0.08 & 0.09 & 0.930\\
\hline
\end{tabular*}
\end{center}
\end{table}
\subsection{Timescale}

It is important to note that these $\langle f_{i(k)}\rangle $ values
are not reached instantaneously: small grains produced after an impact
need time to reach the remote aphelion of their high $a$ and high $e$
orbits. If $dt_{ej(k)}$ is the typical time needed for a small grain
produced in the inner disc to reach $r=10$AU when placed on a high $a_{k}$ and
$e_{k}$ orbit, then during a time step $dt_{1}$, the fraction of
produced $k$ grains that leaves the system will be approximated
through the simplified relation:
\begin{equation}
f'_{i(k)} = (1-f_{i(k)}) . (1-\exp^{-\frac{dt_{1}}{dt_{ej(k)}} })
\label{f'}
\end{equation}
Bodies that did not leave the system during $dt_{1}$ are then added to
a "bodies on their way to leave" subdivision of the $k$ population,
that will in turn decrease by a
$(1-f_{i(k)}).(1-\exp^{-\frac{dt_{2}}{dt_{ej(k)}} })$ fraction at the
next time step $dt_{2}$, etc...

\subsection{Collisional destruction outside the inner disc}

Another possible effect affecting the smallest high $\beta$ particles
is that a fraction of them might be destroyed by collisions $outside$ the
considered inner disc, since they spend an important fraction of their orbit
close to their apoastron which might lie beyond 10\,AU. These
collisions would prevent them from re-entering the inner system.
Such collisions are by definition not modeled by
the collisional evolution Equ.\ref{dn}, and this could lead to an
overestimation of the density of these bodies.

Taking into account this apoastron-collisions removal effect
requires to make physical
assumption on the $external$ parts of the disc and would add several badly
constrained free parameters to our already large set of variables, in
particular the rates and timescales for these collisional destructions as
a function of $\beta$.
We nevertheless tried to investigate the possible
importance of this effect by performing test runs where we
artificially introduced a new parameter
$f_{cl(k)}$, standing for the fraction of $k$ bodies destroyed by collisions
in the $r>10\,$AU region and the corresponding destruction time scale
$dt_{cl(k)}$.
The induced removal of $k$ bodies is then treated the same way as in
Equ.\ref{f'}. $f_{cl(k)}$ might be taken
equal to the fraction of $\beta_{k}$ grains produced within the inner
1--10\,AU disc which have their periastron outside 10\,AU.
The dependency of $dt_{cl(k)}$ with $\beta_{k}$ is more difficult to establish,
and several values will be explored.

As will be shown in section\,4.7, the obtained results do not significantly
depart from our "nominal" case. As a consequence, and for sake of clarity,
we chose to neglect, in a first approximation, this apoastron-collision
effect.

\subsection{Particles with $\beta > 0.5$} 

Even if these particles' ultimate fate is to leave the system, their
ejection takes also a certain amount of time and numerous very small
grains, in this transition phase towards ejection, might be present in
the system and thus collisionaly interacting with other objects.  As
a consequence, our runs will be performed with 2 bins below the
limiting $\beta=0.5$ size. The fraction of $\beta>0.5$ bodies that do
not leave the system after an impact is computed the same way as in
Equ.\ref{f'}, by numerically estimating $dt_{ej(k)}$ and setting
$f_{i(k)}=0$.

\section{Results}
\label{results}

We present here the results obtained for several runs exploring all
important parameters the system's collisional evolution depends on.
As previously mentioned, we define as our "nominal" case the one
defined in section 2.1. Initial conditions for this reference case are
summarized in Tab.\ref{nom}. For sake of clarity, all other parameters
are separately explored in individual runs, even though some
parameters should in principle not be independently explored,
like in particular
the value of $R_{pr}$ (i.e. the grains' porosity) and the
fragmentation and/or cratering prescriptions.  All runs are
carried out until $t_{final}=10^{7}$\,years, i.e. approximately one
third of the minimum age of the system \citep{bar99}.

\begin{table}
\caption[]{Initial parameters for the nominal reference run
(see text for details)}
\label{nom}
\begin{tabular*}{\columnwidth} {ll}
\hline
Minimun size bin  & $3.15\,10^{-4}$cm\\
Maximum size bin  & $5.4\,10^{6}$cm\\
$R_{pr}$ ($\beta=0.5$) & $5\,10^{-4}$cm\\
$\langle e_{0} \rangle$ & 0.1\\
$\langle i_{0} \rangle$ & 0.05\\
$Q_{*}$ law & \citet{benz99}\\
      $b_{1}/b_{2}$ & 1.5\\
      $R_{s(i)}$ & $R_{(i)}/2.10^{5}$ \\    
excavation coefficient $\alpha$ & $10^{-8}$\,s$^{2}$.cm$^{-2}$\\
Initial density & $dN\propto R^{-3.5}dR$ distribution \\
   & with $M_{dust}= 2\,10^{21}$\,g in the annulus\\
 & ($M_{dust}= 2\,10^{22}$\,g in the whole inner disc)\\
\hline
\end{tabular*}
\end{table}

\subsection{Nominal case}
\label{nominal}
\begin{figure}
\includegraphics[angle=0,origin=br,width=\columnwidth]{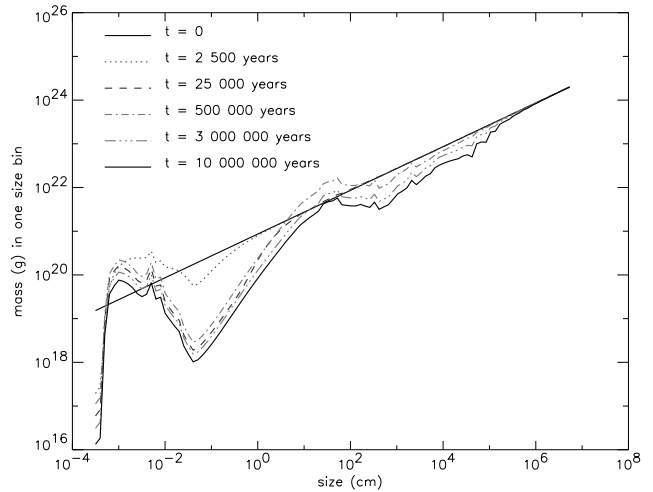}
\caption[]{Size distribution for the low-mass nominal system (see
table\ref{nom}) at 5 different epochs.  Note that the y-axis displays
the mass contained in one size bin, which is a correct way of
displaying the mass distribution since all size bins are equally
spaced in a logarithmic scale. This plot is more "visual" than a more
classical $dN(R)$ one, since it can be directly interpreted in terms
of mass contribution (and mass loss or mass increase) of each size
range to the total mass. }
\label{nom1}
\end{figure}

Fig.~\ref{nom1} shows clearly how the system quickly departs from the
initial $R^{-3.5}$ distribution. A wavy structure rapidly appears
because of the minimum size cut-off, in accordance with \citet{bag94}.
This structure is building up progressively, starting from the lowest
sizes and expanding towards the bigger objects bins. A quasi
steady-state is reached after $\sim10^{6}$\,years and no significant
further evolution of the system is observed in the next
$9\,10^{6}$\,years, except for a slow decrease of the system's total
density.  As could be logically expected, the wavy structure is the
most significant in the small size domain. There is in particular a
strong mass depletion, of a factor $simeq\,40$, in the
$10^{-2}$--$1$\,cm range, with the lowest density point around $R\sim
0.1$\,cm. This depletion has 2 distinct causes: 1) The overabundance
of very small particles due to the size cut-off.  Note however that this
overabundance, though still present, is significantly damped or even
erased for the smallest particles (close to $R_{pr}$),
because these bodies spend a significant fraction of their orbits
outside the inner disc (see the $f_{i(k)}$ parameter in
Tab.\ref{tabf}).  2) The high $\langle dv\rangle$ values for impacts
involving particles close to $R_{pr}$, which are on highly eccentric
orbits.

\begin{figure}
\includegraphics[angle=0,origin=br,width=\columnwidth]{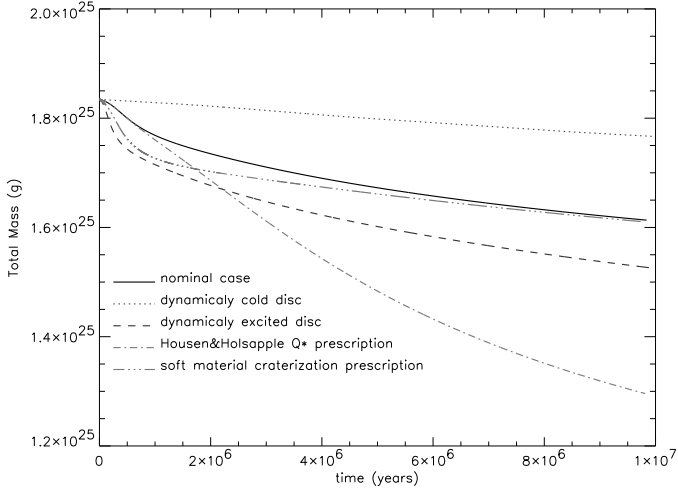}
\caption[]{Temporal evolution of the system's total mass for different cases}
\label{evol}
\end{figure}

Another important result is that the total mass loss of the system
over $10^{7}$\,years remains relatively limited, i.e. less than 12 \%
(cf. Fig.\ref{evol}). Furthermore, the ratio
$M_{dust}/M_{planetesimals}$, after large initial variations,
progressively converges towards a value which is $\simeq\,1/3$ of the
$dN\,=\,CR^{-3.5}dR$ power law value. This is mostly due to a decrease
of $M_{dust}$, with $M_{planetesimals}$ being almost constant.

\begin{figure}
\includegraphics[angle=0,origin=br,width=\columnwidth]{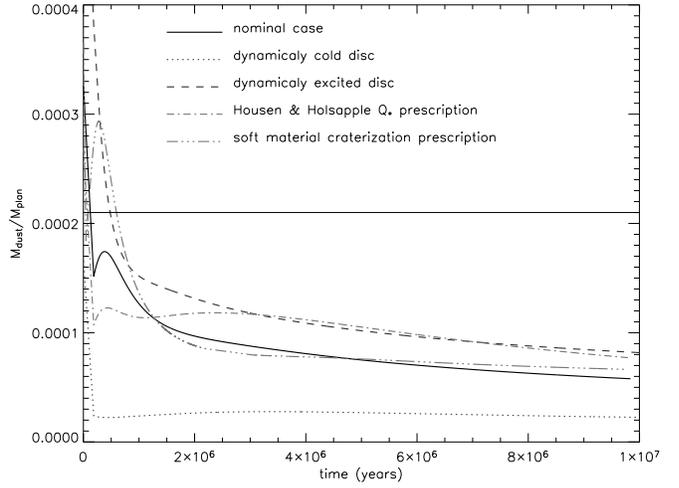}
\caption[]{Temporal evolution of the ratio $M_{dust}/M_{planetesimals}$ for
different cases, where $M_{dust}$ is the total mass of all objects smaller
than 1\,mm and $M_{planetesimals}$ is the mass of all objects bigger
than 10\,km. The horizontal line gives the initial value corresponding
to an academic $dN\,=\,CR^{-3.5}dR$ size distribution.}
\label{evolfrac}
\end{figure}

\subsection{Massive disc} 
\label{massive}
\begin{figure}
\includegraphics[angle=0,origin=br,width=\columnwidth]{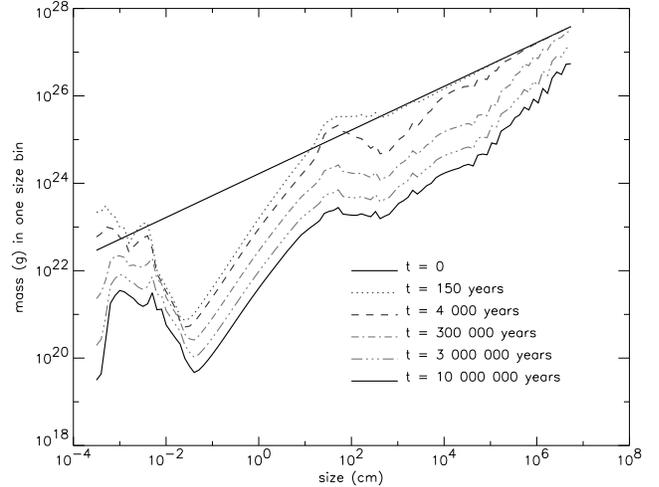}
\caption[]{Size distribution at 5 different epochs for the massive-disc
case, where the total mass of the system is chosen in order to
match the planetesimal mass estimates deduced from FEB mechanism analysis
(see section 2.2).}
\label{mass1}
\end{figure}

\begin{figure}
\includegraphics[angle=0,origin=br,width=\columnwidth]{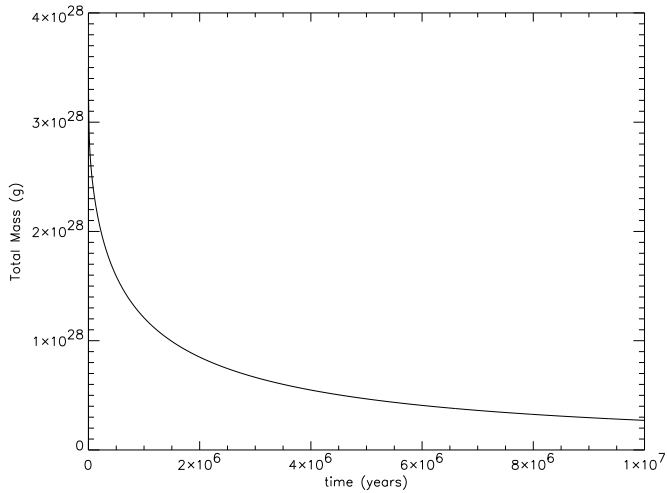}
\caption[]{Temporal evolution of the system's total mass for the
massive disc case}
\label{evolm}
\end{figure}

The massive-disc case turns out to be significantly different
(Fig.\ref{mass1}).  Although a quasi steady-state is here also rapidly
reached and has a profile similar to that of the nominal case, this
steady-state is obtained for a much higher density. This leads to much
faster mass loss than in the previous case, exceeding one order of
magnitude at the end of the simulation (Fig.\ref{evolm}), since mass
loss in a given collisional system increases with the square of the
system's density. We discuss the implications of these results on the
FEB phenomena in section \ref{feb}.

\subsection{Role of the disc's excitation} 
\label{ei}
\begin{figure}
\includegraphics[angle=0,origin=br,width=\columnwidth]{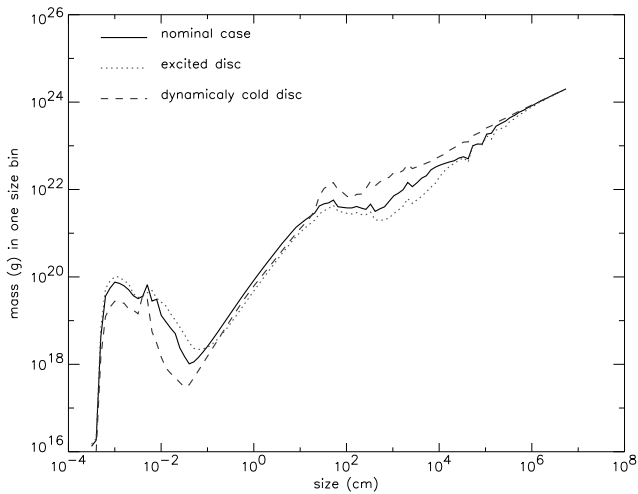}
\caption[]{Final distribution (at $t=10^{7}$years) for different levels
of the disc's dynamical excitation}
\label{compe}
\end{figure}

Apart from the nominal case with $\langle i\rangle=1/2\langle
e\rangle=0.05$, two different disc excitations have been tested: one
low excitation case at $\langle i\rangle=0.0125=1/2\langle e\rangle$
and one high excitation case at $\langle i\rangle=0.1=1/2\langle
e\rangle$, all other parameters being equal (Fig.\ref{compe}).

As could logically be expected, the total mass loss is much higher in the
high excitation case, $\simeq\,18\%$, than in the low excitation case,
$\simeq\,4\%$. Nevertheless, the steady-state regime profile is
significantly different for both runs. The density well in the
0.01-1\,cm range is in particular much deeper for the dynamically cold
disc. This is a fully logical result when considering the fact that
the radiation pressure induced high $e$ of the smallest grains only
weakly depends on the dynamical state of the parent bodies (Equ.\ref{pre}).
As a consequence, the contrast between the excitation, and thus the
shattering power, of the smallest grains and
that of the rest of the particles is very high. The destruction rate of
bodies in the 0.01-1\,cm range by small grains is thus at the same level
than in the nominal case, whereas the production rate of 0.01-1\,cm grains
by collisions between bigger objects is much lower, hence the deeper
density well.
Conversely, for the high excitation case, the contrast between the
small grains' and bigger objects' destructive powers is significantly damped,
hence a shallower density drop in the 0.01-1cm range.

\subsection{Porosity, value of $R_{pr}$} 
\label{porosite}
\begin{figure}
\includegraphics[angle=0,origin=br,width=\columnwidth]{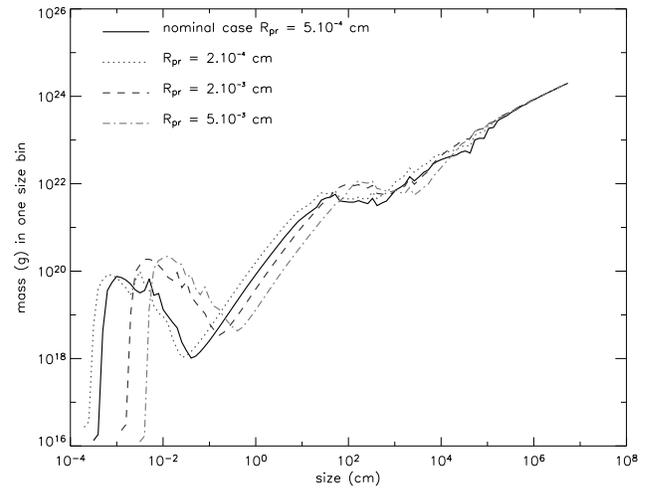}
\caption[]{Final distribution (at $t=10^{7}$years) for different values
of $R_{pr}$}
\label{comphr}
\end{figure}

As previously discussed, our nominal case corresponds to compact low
porosity grains and $R_{pr}=5\,\mu m$. We investigated different
porosities, and thus different $R_{pr}$ values, all other parameters
being equal (with always 2 size boxes below $R_{pr}$).  As can be
clearly seen in Fig.\ref{comphr}, changing the value of $R_{pr}$
results mainly in shifting the wavy structure without affecting its
overall profile. Although it is not strictly speaking a homothetic
shift, mainly because of the complexity of the $Q_{*}$ prescription,
differences are minor ones, and the density drop is always located at
roughly $100\,R_{pr}$.

\subsection{$Q_{*}$ prescription}
\label{Qstar}
\begin{figure}
\includegraphics[angle=0,origin=br,width=\columnwidth]{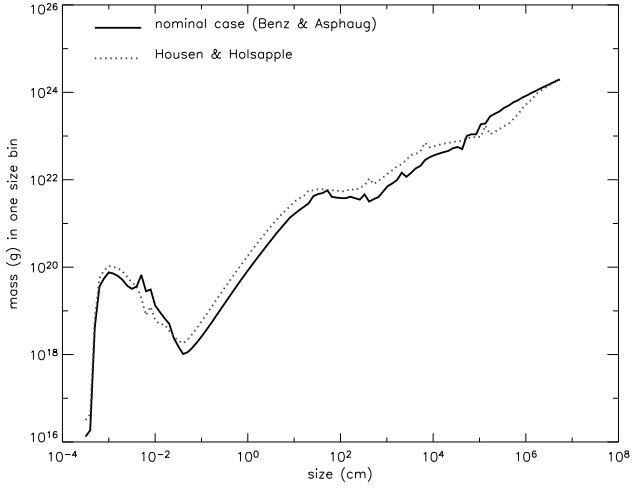}
\caption[]{Final mass distribution (at $t=10^{7}$\,years) for different
$Q_{*}$ prescriptions}
\label{qcomp}
\end{figure}

Taking the Housen \& Holsapple (1990) and Holsapple (1994) prescription
for $Q_{*}$ leads to a final size distribution which is remarkably
close to the nominal Benz \& Asphaug (1999) case (Fig.\ref{qcomp}).
The main difference is a more defined density drop for objects bigger
than $10^{5}$\,cm, which is directly due to the fact that $Q_{*}$
values for bodies in this size range are significantly lower than in
the Benz \& Asphaug (1999) model. This deficit in large bodies is the
reason for the more significant total mass loss in the system
(Fig.\ref{evol}), since these bodies contain most of the system's
mass.  Note however that this total mass loss does not reach very
large values, remaining limited to less than $30\,\%$, and that the
global dust to planetesimals mass ratio is also relatively close to
the nominal case value (Fig.\ref{evolfrac}).

\subsection{Fragmentation and Cratering prescriptions}
\label{fragcrat}
\begin{figure}
\includegraphics[angle=0,origin=br,width=\columnwidth]{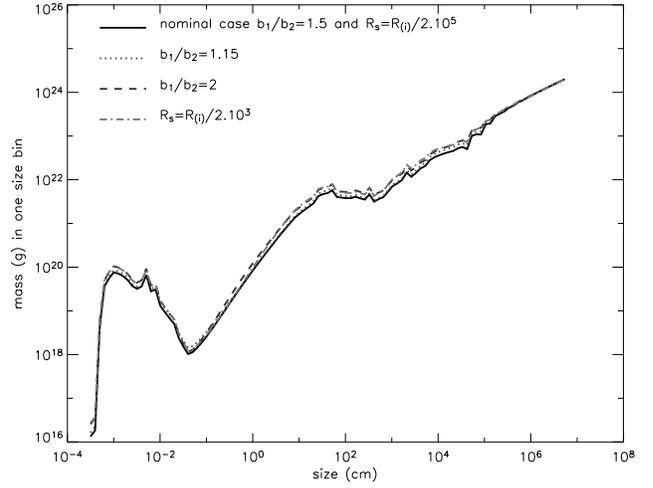}
\caption[]{Final mass distribution (at $t=10^{7}$\,years) 
for different values of the free parameters of our bimodal power law:
i.e. the ratio of their slopes $q_{1}/q_{2}$ and the position of the slope
changing size $R_{s}$ with respect to the size $R_{i}$ of the impacted body. }
\label{frcomp}
\end{figure}

As previously shown, the main free parameters for our fragmentation
prescription are the ratio $q_{1}/q_{2}$ and the size $R_{s}$ of the
slope transition. These parameters were both explored in independent
runs whose results are presented in Fig(\ref{frcomp}). As can be
clearly seen, the values of $q_{1}/q_{2}$ and $R_{s}$ only moderately
affect the final size distribution within the system.

\begin{figure}
\includegraphics[angle=0,origin=br,width=\columnwidth]{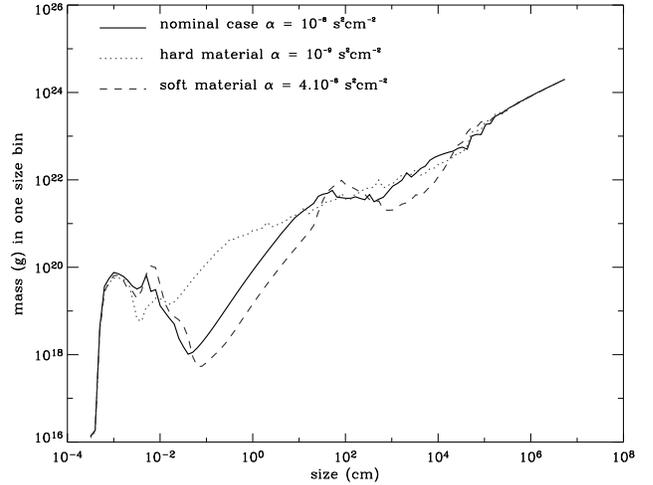}
\caption[]{Final mass distribution (at $t=10^{7}$\,years) for different
values of the excavation coefficient $\alpha$}
\label{crcomp}
\end{figure}

As appears in Fig(\ref{crcomp}), the cratering prescription, in
particular the value of the excavation coefficient $\alpha$, has a
more significant effect on the physical evolution of the system.
Taking a very hard material prescription ($\alpha\,=\,10^{-9}$) leads
indeed to a final size distribution which is very close to a
$dN\,=\,CR^{-3.5}dR$ power law. The density drop in the sub-centimeter
size-range is in particular significantly reduced, with only a
factor 8 drop in a narrow region around
$10^{-2}\,$cm. Conversely, the very soft material run
($\alpha\,=\,4.10^{-8}$) leads to a deeper density drop and a more
pronounced wavy structure throughout the size distribution.  This
dependency of the size distribution profile on the cratering
prescription is easily understandable when realizing that, in the
0.01 to 1\,cm domain, cratering is a much more efficient process
than fragmentation in terms of mass removal (Tab.\ref{fracra}); mainly
because of the cratering events due to the high $e$ grains in the
$R_{pr}$ to $\simeq\,10\,R_{pr}$ range.
Note however, that for the system as a whole, it is fragmentation which is
clearly the dominating mass removing  process (Tab.\ref{fracra}).

\begin{table}
\caption[]{This table sums up, for all objects within 3 different
size ranges, the respective amount of mass that is removed by all
cratering and all fragmenting impacts. These values are obtained in
the steady state regime for the nominal case.}
\label{fracra}
\begin{tabular*}{\columnwidth}{lcccc}
\hline
size range (cm)& R$<$$5.\,10^{-3}$ & $0.01$$<$R$<$1 & $10^{5}$$<$R & total\\
\hline
fraction of & & & & \\
mass removed: & & & & \\
by fragmentation & 0.89 & 0.13 & 0.76 & 0.67 \\
by cratering & 0.11 & 0.87 & 0.24 & 0.33 \\
\hline
\end{tabular*}
\end{table}

\subsection{Collisions outside the inner disc}

\begin{figure}
\includegraphics[angle=0,origin=br,width=\columnwidth]{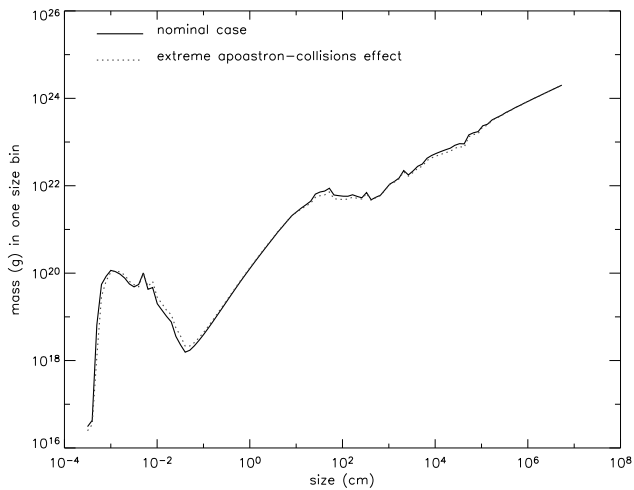}
\caption[]{Final mass distribution (at $t=10^{7}$\,years) for an academic
case with a very efficient removal of small particles by hypothetic collisions
in the $r>10\,$AU region (see text for details).}
\label{collout}
\end{figure}

As discussed in section\,3.2, we chose to perform additional test runs
checking the possible influence of small particles removal by collisions
outside the inner disc. This effect is arbitrarily parameterised by the two
parameters $dt_{cl(k)}$ and $f_{cl(k)}$ (see section\,3.2).

We present here results obtained for the most extreme case, where
$dt_{cl(k)}$ was unrealistically supposed to be equal to one orbital period of
a $\beta_{k}$ particle. As appears clearly from Fig.\ref{collout},
differences with the nominal case remain marginal. As expected, the main
difference is found for bins just below the $\beta=0.5$ cut-off, with a
factor 4 number density difference for the first bin corresponding to
bound orbits ($\beta_{k}=0.49$). Nevertheless, this difference already drops
to $25\%$ for particles of size $2R_{pr}$ ($\beta_{k}=0.24$).
As a consequence, the sharp density drop induced by the apoastron collisions
effect remains confined to a narrow size range of particles. Furthermore,
this density drop does not have significant consequences on the rest
of the size distribution and doesn't affect the global profile of the wavy
size distribution. This is because it affects particles which are already
strongly depleted because of their high $a$ and $e$ (low $f_{i(k)}$) values.
Besides, this removing effect's dependency on $\beta$ is relatively
similar to that of the one induced by low $f_{i(k)}$) values; it will thus only
tend to reinforce an effect already taken into account. 
Thus, further depleting these populations does not lead to drastic changes.

\section{Discussion}
\label{discussion}

\subsection{The massive disc case. Problems with the FEB scenario}
\label{feb}
One of the most obvious and easily understandable results is the
strong mass loss for the massive disc case. As previously stated, the
system loses more than 90\% of its total mass over $10^{7}\,$years.
The problem gets even worse when trying to extrapolate this mass loss
to the past.  This might be done when noticing that, apart from the
initial transition phase, the system's mass loss in the steady state
regime might be fitted by a $M_{(t)}=(a+b.t)^{-1}$ law (which is a
logical result since the mass loss is proportional to the square of
the system's total mass), with $a=5.2\,10^{-29}\,g^{-1}$ and
$b=3.2\,10^{-35}\,g^{-1}s^{-1}$. It is easy to see that
extrapolating this law to the past
leads to masses that become rapidly unrealistically high.  Thus, the
planetesimal density required to sustain the FEB phenomenon
corresponds to a very rapidly evolving system and cannot be maintained
over a long period of time.  One could argue that considering a
dynamically colder system would significantly reduce the system's mass
loss.  But this would not solve the problem, since the strength of the
FEB producing mean-motion resonances directly depends on the system's
excitation, so that reducing the disc's excitation would require an
even higher number density of planetesimals in order to get the
observed FEB rate \citep{thebeu01}.

The problem with the sofar accepted FEB scenario is then the
following: from combined observations and modelling, the massive disc
required to sustain the observed activity should erode significantly
within less than $10^{6}$yrs, giving a natural end to the FEB
phenomenon. If this was to be the case, then we would be presently
witnessing a very transient phenomenon. This does not appear
satisfactory from a statistical point of view.  Should this mean that
the FEB scenario should be rejected as a whole ? We believe that it is
too early to state anything definitely. There are several reasons
for that:

We must first recall that the estimate for the necessary disc
population for sustaining the FEB activity is derived through a chain
calculation which depends on several poorly constrained parameters
\citep[see the extended discussion in ][]{thebeu01}. \citet{thebeu01}
(Eqs. 7 and 8) showed that the most crucial parameter is here
$R_{\mathrm{FEB}}$, i.e.  the minimum size of bodies able to become
observable FEBs, since the deduced total mass scales roughly as
$R_{\mathrm{FEB}}^{q-1}$ in the simplified case where a power law of
index $q$ applies for the size distribution above $R_{\mathrm{FEB}}$,
so any change to $R_{\mathrm{FEB}}$ may induce drastic changes to the
estimated disc mass.  \citet{thebeu01} assumed
$R_{\mathrm{FEB}}=15\,$km, but this value is poorly known and could
easily vary by one order of magnitude.  $R_{\mathrm{FEB}}$ exists
because bodies smaller than $R_{\mathrm{FEB}}$ are assumed to
evaporate too quickly and consequently make too few periastron
passages in the refractory evaporation zone ($\la 0.4\,$AU) to
significantly contribute to the observable spectral activity. The
value of $R_{\mathrm{FEB}}$ is thus related to the evaporation rate of
the FEBs themselves. Simulations of the dynamics of the material
produced by FEB evaporation (Beust et al. 1996) led to derive
production rates of a few $10^7\mbox{kg\,s}^{-1}$ as necessary to
yield observable spectral components. We believe that this part of the
scenario needs to be revised. The main reason for that is that in
Beust et al. (1996) simulations, the material escaped from the
FEBs was assumed for simplicity to consist of the metallic ions we
study and volatile material.  The metallic ions undergo a strong
radiation pressure from the star while this is not the case for the
volatiles. Hence the volatiles retain the metallic ions around the FEB
coma for a while, leading to observable components. The production
rate was then derived from the necessary amount of volatiles to retain
the ions, and from assumptions about the chemical composition of the
body. All this is obviously the weakest part of the scenario.

Besides, \citet{kar01} showed that if the FEB
progenitors are supposed to originate from 4--5 AU from the star,
they should no longer contain ices today (i.e. volatile material),
apart from an eventual residual core. More recently, \citet{kar03}
made an independent theoretical study of the evaporation
behaviour of such objects when they gradually approach the star on
repeated periastron passages. The evaporation rates derived are thus
independent from any observation. Basically, this work shows that
$\sim 10\,$km sized bodies fully evaporate with repeated
periastron passages, and that evaporation rates of a few
$10^7\mbox{kg\,s}^{-1}$ are actually reached, but this occurs
only when the periastron is less than $\sim 0.2$\,AU, i.e. well
inside the dust evaporation zone and shortly before the final
evaporation of the body. Before that, any FEB entering the dust evaporation
zone ($\la 0.4\,$AU) but for which the periastron has not yet reached
$0.2$\,AU actually evaporates, but at a weaker rate. If it is small,
it thus survives more periastron passages than in previous estimates,
and may contribute to the observational statistics.

However, whether bodies with no or very few volatiles may generate 
observable components is questionable, as volatiles have a crucial role
in the dynamics of the metallic ions. Within the refractory
material, some species suffering low radiation pressure, and that
are probably abundant (Carbon, Silicon,\ldots) may play the retaining
role of volatiles. Obviously this question must be reinvestigated 
with more realistic simulations, but a probably outcome will be that
$R_{\mathrm{FEB}}$ could end up to be at least one order of magnitude
less than previously estimated. In this context, our chain calculation
would lead to a much lower disc mass necessary for sustaining the FEB
activity.

In this context, it is impossible to rule out the FEB scenario
on this basis alone, but this remain a problematic possibility.
All we can presently state is that the disc populations
inferred by \citet{thebeu01} are unrealistic and that the 
FEB scenario should at least be reinvestigated much more carefully.

\subsection{Departure from the $R^{-3.5}$ profile}

Putting aside the peculiar massive disc problem, the most striking
result, present for almost all tested simulations, is a final
size-distribution that significantly departs from a $R^{-3.5}$ power
law, especially in the small size domain. The only exception to this
behaviour is a run with a very hard material parameter for the
cratering prescription, which means that alternative
size-distribution profiles cannot be completely ruled out, although
they seem to represent a marginal possibility.  Of course, due to the
complexity of the studied problem, all free parameters could not be
exhaustively explored. Besides, there are some parameters that are
strongly coupled, i.e. fragmentation and cratering prescriptions
should in principle not be independently explored.  Nevertheless,
there seems to be a global tendency towards a common feature which
consists of a lack of objects close to the limiting ejection size
$R_{pr}$, a density peak at $\simeq\,2\,R_{pr}$ and a sharp density
drop compared to the $R^{-3.5}$ law, of one or two orders of
magnitude in mass, at $\simeq\,100\,R_{pr}$. It is also important to
note that a very badly constrained parameter such as the disc's
dynamical excitation does not seem to have a crucial influence on the
profile, thus reinforcing the genericity of this result.

These departures from the $dN\,=\,CR^{-3.5}dR$ power law do not lead
to radical changes in the global dust vs. planetesimal mass ratio in
the system, which only decreases by a factor $\simeq$3--4. If
$2\,10^{22}\,$g is a typical value for the amount of dust (i.e. $R<1\,$mm)
in the inner 10 AU region (see section I), then we estimate from
our results that the corresponding mass
of $1\,$km$<$R$<50\,$km objects should be $\simeq$
3.5--7$\,10^{-2}\,M_{\oplus}$, which remains a value comparable
to the one roughly derived from a $R^{-3.5}$ power law (see section I).
Even stretching this value up to the
$1\,$km$<$R$<500$\,km range does not lead to more than
$0.15\,M_{\oplus}$ of "large" objects.  Our calculations thus
quantitatively confirm what had been previously inferred from
questionable assumptions (an $R^{-3.5}$ power law): there $is$ a lack
of objects, that holds even for large planetesimals, in the inner
disc.

This is an additional problem for the FEB scenario, since
this value is far from being enough in order to account for the sharp
incompatibility between the amount of observed dust and the required
amount of FEB inducing planetesimals.
In any case, it appears clearly that the FEB model as it
is currently accepted cannot be compatible with a "reasonable" estimate
of the dust production rate in the inner disc.

\subsection{Collisional erosion vs. cometary evaporation}

Let us recall that the precise SED fit performed by \citet{ligre98}
was obtained assuming that the dust is of pure cometary origin and is
not affected by collision processes.  On the contrary, in our
simulations we implicitly made the assumption that the inner \bp\ dust
disc is made of collisional debris. We do believe that our results
retrospectively justify this assumption, although without ruling out
the possible presence of evaporating bodies, and this for several
reasons.

\begin{figure}
\includegraphics[angle=0,origin=br,width=\columnwidth]{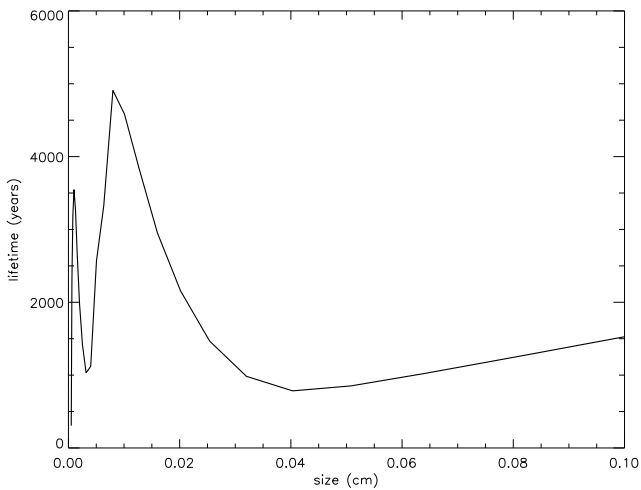}
\caption[]{Typical lifetime, as a function of its size, of a dust
particle in the inner disc before destruction by collision (steady
state regime of the nominal case)}
\label{lifetime}
\end{figure}

\begin{enumerate}
\item For what is presently known about the inner disc, the collision
production hypothesis seems to be quantitatively more $generic$ than
the concurrent cometary evaporation. In his Equ.2 \citet{lec98}
proposed a simple expression in order to estimate the number $N_{co}$
of currently evaporating comets in a dust disc, where $N_{co}$ directly
depends on $M_{dust}$, the total dust mass, and $t^{-1}_{dust}$, the
typical lifetime for a dust particle before destruction.  Taking the
same Halley-at-1-AU evaporation rates as \citet{lec98} and bodies of
20\,km in radius leads to
$N_{co}\simeq\,1.5\,10^{4}*(t_{dust}/10^{4}\,yrs)^{-1}$.  Considering
that collisional lifetimes of dust grains are of the order of
$10^{3}\,$years in the inner Beta-Pic disc (Fig.\ref{lifetime}), we
get $\simeq\,2.6\,10^{-3}\,M_{\oplus}$ of evaporating comets in the
$r<10\,$AU region. This value represents $\simeq\,$10\% of the
estimated total mass of kilometre-sized objects
(cf. Tab.\ref{summary}), which should mean that one out of ten
planetesimal objects is an evaporating comet. Although such a
possibility cannot be completely ruled out, it seems nevertheless
rather unlikely since \citet{kar01} has showed that all kilometre-sized
objects originating from the 5 AU region should no longer contain volatile
material today.
One could argue that the previous calculation depends
on poorly constrained parameters and that the requested number of
evaporating bodies could be lower. But even in this case there is one
crucial problem to solve : what is the mechanism constantly refilling
the inner disc with so many fresh comets?
The main difficulty is that this refilling has to be very effective and
rapid, since the average time needed for a 10\,km body at 5 AU to lose all
its volatile material is less than 100\,years (see Fig.2 of
Karmann et al.; 2001).

\item A more conclusive argument is that the present simulations show
that the presence of $\simeq\,2\,10^{22}g$ of dust in the $<10\,$AU
region might be explained, within a moderate mass disc, by collisional
processes $alone$. Moreover, such a low mass disc is consistent with
what should be expected, in the inner disc, considering the age of the
system: a disc of debris left over after the accretion process of
planetary embryos (see next subsection). In short, there is no $need$
for a cometary activity in terms of production of the observed dust.
\item In any case, even if all the dust was to be produced by
evaporating comets, then the present simulations show that mutual
collisions within such a $\simeq\,2\,10^{22}$\,g dust disc would anyway
be $unavoidable$. This would strongly affect the size distribution,
which would probably tend towards the steady-state profiles displayed
in section 4.
\end{enumerate}

Of course, these arguments are relevant only for the $inner$
$\beta$\,Pic disc. We do not rule out the possibility that comet
evaporation could be a dominant dust-production source in the outer
parts of the system, as suggested by the Orbital Evaporating Bodies
scenario proposed by \citet{lec98} for the region beyond 70\,AU.

This might appear to be a somehow paradoxal result, since evaporation
processes should be more effective in the inner regions. But let
us once again stress that these higher evaporation rates are
precisely what makes it difficult to find a way to sustain evaporation
activity over long time scales in the inner disc (as previously
mentioned, a 10\,km object evaporates in less than 100 years at 5\,AU).
At larger distances, volatile evaporation rates are much lower,
thus reducing the crucial problem of "refilling" the evaporation
region with fresh material. Of course, distances from the star must
remain within the limiting distance at which evaporation is possible
, i.e. 100--150\,AU for CO \citep{lec96}.

\subsection{Presence of already formed planetary embryos}

Of course, one cannot rule out the possible presence of isolated much
more massive objects, such as planets or planetary embryos, whose
isolation decouples them from the collisional cascade responsible for
the dust production (a possibility considered by \citet{wya02}
for the Fomalhaut system). In fact, the low mass in the
dust-to-planetesimals range could be interpreted as the consequence
of the presence of such planetary embryos: most of the initial mass of
the system would already have been accreted in these embryos, leaving
a sparse disc of remnants.  This would be in accordance with the
estimated age of the system, a few $10^{7}$\,years, which
significantly exceeds the expected timespan for the formation of
planetary embryos \citep[e.g.][]{lis93}, so that $if$ embryos have to
form, then they should be already here.

Another argument previously proposed in favour of the presence
of already formed massive
embryos is that such objects are a good way to explain the disc's
thickness. \citet{arty97} estimated that numerous Moon-sized bodies
are required in order to induce vertical velocity dispersions of
smaller bodies of the order of $0.1\,v_{kep}$. Nevertheless, as
pointed out by \citet{mou97}, a giant planet on a slightly inclined
orbit (like the planet required to explain the warp in the outer
regions) could achieve just the same result: rapid precession of the
dust particles orbits in the inner regions would lead to thicken the
disc so that the aspect ratio appears to be equal to the planet's
inclination.

\section{Conclusions and Perspectives: towards a coherent picture of
the inner $\beta$ Pictoris disc?}
\label{conclusion}

The present study show that the observed $10^{21}\,$to$\,10^{22}\,$g
of dust in the inner disc is compatible with what would be expected
from a collisional cascade within a disc of a few
$10^{-2}\,M_{\oplus}$ bodies ranging from micron to
kilometre-sized objects, without the need for any additional
cometary-evaporation activity.  And even if there was such a cometary
activity, the required density of objects would inevitably lead to
important collisional effects.

Simulations also show that the size distribution settles towards a
quasi-equilibrium state that strongly departs from the classical
$dN\propto\,R^{-3.5}dR$ Dohnanyi power law. This is particularly true
for the smallest grains close to the radiation pressure ejection
limit.  However, these departures do not too strongly affect the
global Dust-to-Planetesimal mass ratio and cannot account for the
incompatibility between the small amount of observed dust and the huge
number of kilometre-sized FEBs requested to sustain the transient
absorption features activity. Furthermore, our runs show that this
requested mass of FEBs leads to a much too rapidly collisionaly
eroding disc that cannot survive on long timescales. The FEB scenario
thus cannot hold in its present form and has to be seriously revised.

We might thus converge towards a coherent picture of the inner $\beta$
Pictoris disc: This inner disc should be boundered by one giant planet
of $\simeq\,1\,M_{Jup}$, located around 10\,AU on a slightly inclined
(in order to explain the observed outer warp) and possibly eccentric
orbit (in order to trigger the FEB activity).  The observed amount of
dust should be produced by collisional erosion within a low mass
disc. Such a low mass disc could be made of debris leftover after the
accretion of one or several planetary embryos, the presence of which
is fully compatible with the estimated age of the system, i.e. a few
$10^{7}$\,years.
In other words, we should be now witnessing a planetary system in a
late or at least intermediate stage. The bulk of the accretion process
is over, but a consequent disc of remnants is still present and
collisionaly eroding.

Our results would also help putting new constraints on the SED fits
that are usually performed to derive dust densities and radial
distributions from observed spectra. Let us recall that the dust mass
estimations for the inner disc, which we used as inputs for our simulations,
have been computed either by
postulating that grains are of cometary origin \citep{ligre98} or by
doing a pure mathematical fit with several free parameters
\citep{pan97}.  In this respect, it would be interesting to perform a
work similar to that of \citet{ligre98} but with a population of
collisionaly produced grains as input. 
An interesting attempt at doing such a kind of study has been recently
made by \citet{wya02} in their very detailed study of the Fomalhaut's
debris disc. Nevertheless, their precise fit of the SED was made
assuming a single power law for the size distribution (even though the authors
were fully aware of the fact that such an academic distribution cannot
hold for the smallest grains because of the cutoff effect).
Such an SED-fit analysis goes
beyond the scope of the present paper and requires additional work.

It requires in
particular to model the $whole$ \bp\ disc and not only the
innermost parts that only partially contribute to the total flux.
Only then could the obtained size distribution be compared to
SEDs integrated over the whole disc. A crucial problem would probably
be to see if the underabundance of millimetre-sized objects that we
obtained in the inner disc is also to be found for the system as
a whole; this would then contradict previous estimates stating
that the observed mass of millimetre objects is in accordance
with a $-3.5$ equilibrium power law \citep{arty97}.
Such a study should of course also address more deeply the question
of the physical nature of the dust grains.
It will be the purpose of a forthcoming paper.

\begin{acknowledgements}

J.C. Augereau was supported by a CNES grant and a European Research
Training Network ``The Origin of Planetary Systems'' (PLANETS,
contract number HPRN-CT-2002-00308) fellowship.

\end{acknowledgements}
{}
\clearpage

\end{document}